\documentclass[letter]{aa}
\usepackage{graphicx,epsfig,fancyhdr,psfig,rotate,amsmath,longtable}
  \newcommand{\eg}{\mbox{e.g.}}                          
  \def\simge{\mathrel{\raise1.16pt\hbox{$>$}\kern-7.0pt
    \lower3.06pt\hbox{{$\scriptstyle \sim$}}}}           
  \def\simle{\mathrel{\raise1.16pt\hbox{$<$}\kern-7.0pt
    \lower3.06pt\hbox{{$\scriptstyle \sim$}}}}           

\begin{document}

      \title{Low polarized emission from the core of coronal mass ejections}

      \subtitle{}

      \author{M. Mierla\inst{1,2,3} \and I. Chifu\inst{4,5} \and B. Inhester\inst{4} \and L. Rodriguez\inst{1} \and A. Zhukov\inst{1,6} }


    \offprints{M. M.\\
    \email{marilena@geodin.ro}}
   
   \institute{Solar-Terrestrial Center of Excellence - SIDC, 
               Royal Observatory of Belgium, Brussels, Belgium \and 
               Institute of Geodynamics of the Romanian Academy,
               Bucharest, Romania, RO-020032 \and
               Research Center for Atomic Physics and Astrophysics,
               Faculty of Physics, University of Bucharest, Romania \and
               Max-Planck-Institut f\"{u}r Sonnensystemforschung,
               Katlenburg-Lindau, Germany \and Astronomical Institute of
               the Romanian Academy, Bucharest, Romania \and Skobeltsyn
                Institute of Nuclear Physics, Moscow State University, Russia}
    \date{Published in: Astronomy and Astrophysics, vol 530 (June 2011), L1,
              DOI: 10.1051/0004-6361/201016295}
\abstract {}
{In white-light coronagraph images, cool prominence material is
sometimes observed as bright patches in the core of coronal mass
ejections (CMEs). If, as generally assumed, this emission is caused by
Thomson-scattered light from the solar surface, it should be
strongly polarised tangentially to the solar limb. However, the
observations of a CME made with the SECCHI/STEREO coronagraphs on 31 August 2007 show that the emission from these bright core patches is exceptionally low polarised.} {We used the polarisation ratio method of Moran and Davila (2004) to localise the barycentre of the CME cloud. By analysing the data from both STEREO spacecraft we could resolve the plane-of-the-sky ambiguity this method usually suffers from. Stereoscopic triangulation was used to independently localise the low-polarisation patch relative to
the cloud.}
{We demonstrated for the first time that the bright core material is located close to the centre of the CME cloud. We show that the major part of the CME core emission, more than 85\% in our case, is H$\alpha$ radiation and only a small fraction is Thomson-scattered light. Recent calculations also imply that the plasma density in the patch is 8 10$^8$ cm$^{-3}$ or more
compared to 2.6 10$^6$ cm$^{-3}$ for the Thomson-scattering CME environment
surrounding the core material.}
{}

   \titlerunning{Low polarised emission}

   \authorrunning{Mierla et al.}

   \maketitle


\section{Introduction}

The core of coronal mass ejection (CME) clouds occasionally exposes
very bright concentrated patches in white-light coronagraphs. 
They are interpreted as cool plasma material from a prominence
that was embedded inside the streamer environment of the CME before the eruption. During the eruption process, the prominence is
then expelled along with the surrounding streamer plasma.

Poland and Munro (1976) report one such observation
made on 21 August 1973, 15:11 UT with the Skylab white-light coronagraph and its HeII 30.4 nm spectroheliograph.
About 18 min before, an H$\alpha$ image taken at the Sacramento
Peak Observatory had shown bright patches extending
out to 1.42 R$_{\odot}$ but fading in intensity with time.
Even though the coronagraph field-of-view was limited to above 1.5 R$_{\odot}$, it was concluded that H$\alpha$ radiation contributed to the white-light image because its signal was less polarised in some bright patches than in the surrounding region.

H$\alpha$ radiation is the result of the electronic $j$=3 $\rightarrow$ $j$=2
transition of the hydrogen atom. In equilibrium at an electron temperature
below 50,000~K, the $j$=3 level is populated much more by absorption of the
ambient Ly$\beta$ radiation than by absorption of photospheric H$\alpha$.
This causes a substantial decrease in polarisation of the emitted H$\alpha$
radiation below the theoretical maximum value of 30\% for pure resonant
scattering (Poland and Munro 1976). Besides, the Hanle effect caused by the
coronal magnetic field (Sahal-Brechot et al. 1977, Heinzel et al. 1996) and
collisional depolarisation (Bommier et al. 1986) reduce the amount of
polarisation even further. As a result, the linear polarisation of H$\alpha$
radiation observed in prominences well above the limb ranges from a fraction of
a percent (Gandorfer, 2000; Wiehr and Bianda, 2003) to a few percent (Leroy et
al., 1984).

The white-light emission of the solar K-corona originates in Thomson-scattering of photospheric light by free electrons. Detailed
description of the Thomson-scattering theory can be found in various papers
(\eg, Minnaert 1930; van de Hulst 1950; Billings 1966). The anisotropy
of the incident light causes the observed scattered radiation to exhibit
a polarisation parallel to the visible limb. The degree of polarisation
depends on the distance from the solar surface and on the scattering angle
to the observer. 
Indeed, it has been proposed to use the observed degree of polarisation
to estimate the distance of the coronal scattering volume off the plane
of the sky (POS; \eg, Moran and Davila 2004, Dere et al. 2005,
Vourlidas and Howard 2006). Hence, a reduction of polarisation of a
white-light signal from the corona can in principle also be explained 
by a geometric effect, and Poland and Munro's
conclusion only holds if it is assumed that the H$\alpha$ material is
well embedded in the CME cloud of enhanced electron density and is
located close to the POS of the observer.

    \begin{figure*}[t]
    \centering
    \includegraphics[width=0.9\textwidth,type=eps,ext=.eps,read=.eps]{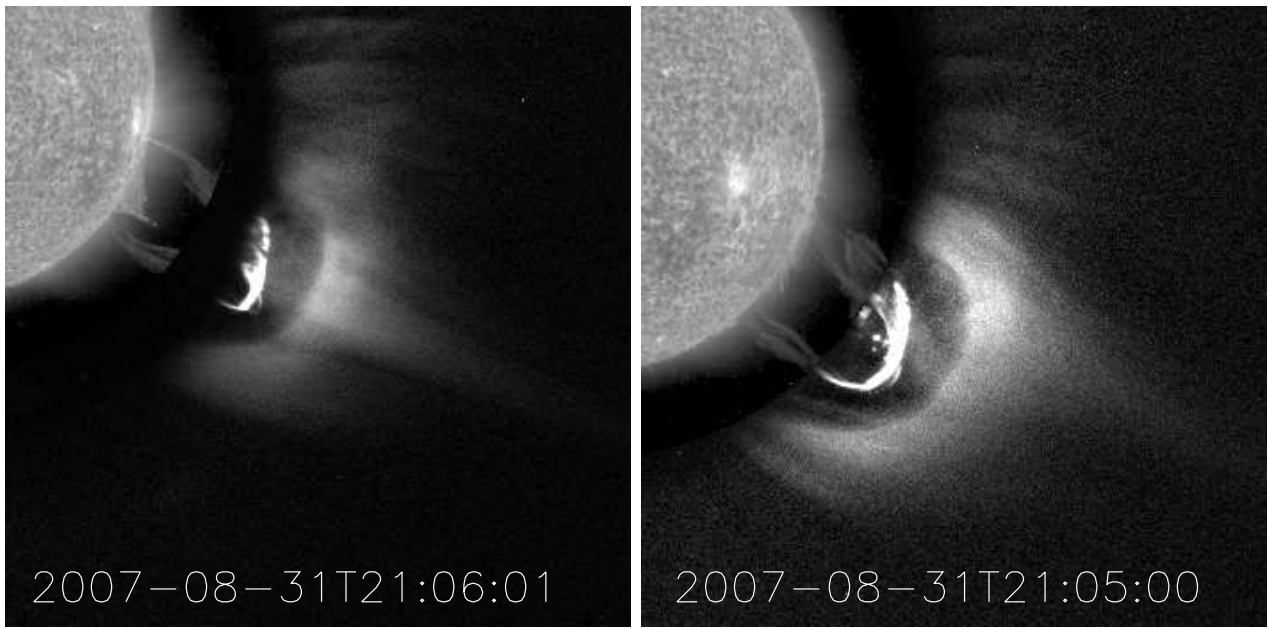}
    \caption{Combined images of EUVI and COR1 on board STEREO A and B. Both A and B images
    are synchronous with respect to the time on the Sun.}
    \label{plotsmin11aug}
    \end{figure*}

In this letter, we report the observation of a similar incidence with
the COR1 coronagraph on board the two STEREO spacecraft A and B. In addition
to the polarisation measurements from which we determine the azimuthal
barycentre position of the CME plasma, a stereoscopic triangulation of
the low-polarisation patch proves that the prominence material is well
embedded inside the CME.

\section{Observations}

EUVI and COR1 are the multi-wavelength EUV telescope and the innermost
coronagraph of the Sun Earth Connection Coronal and Heliospheric
Investigation (SECCHI) instrument suite (Howard et al. 2008) aboard
the twin Solar Terrestrial Relations Observatory spacecraft (STEREO,
see Kaiser et al. 2008). Each of the COR1/STEREO telescopes has a field of view from 1.4 to 4 R$_\odot$
and observes in a white-light waveband 22.5~nm wide centred at the H$\alpha$
line at 656~nm (Thompson and Reginald 2008). The COR1
coronagraphs take polarised images at three different polarisation angles
at 0, 120, and 240 degrees. These primary data allow to derive Stokes I,U and Q
components and finally total ($tB$), polarised ($pB$) 
and unpolarised brightness ($uB=tB-pB$) images. 

The start of a prominence eruption was observed in EUVI HeII bandpass
images at 30.4 nm on 31 August 2007, at around 19:00 UT. Approximately eight hours
before, a dark filament was detected about 100,000 km south of an active region
close to the west limb in the H$\alpha$ images of Kanzelh\"ohe Observatory, which had disappeared the following day
(http://cesar.kso.ac.at/halpha2k/archive/2007).
At about 21:00 UT, the HeII prominence had risen to 1.5 R$_\odot$
and appeared co-spatial with the bright core of a structured CME detected in 
COR1 images (Fig.~\ref{plotsmin11aug}, see for e.g. Cremades
and Bothmer 2004 for the definition of a structured CME).

Preliminary COR1 images of the event studied here revealed patches of 
extremely low polarisation in the bright CME core. These patches 
of low polarisation could faintly be detected even out to about 
7 R$_\odot$ in COR2 at about 2:00 UT the next day. COR2 is the outer coronagraph 
of STEREO/SECCHI, which covers distances from 3 to 15 R$_\odot$. The
outward velocity of these patches started from about 170 km/s in COR1 and
accelerated to 240 km/s in the COR2 field of view.

\section{Data analysis}

\begin{figure}
    \centering
    \includegraphics[width=.48\textwidth,type=eps,ext=.eps,read=.eps]{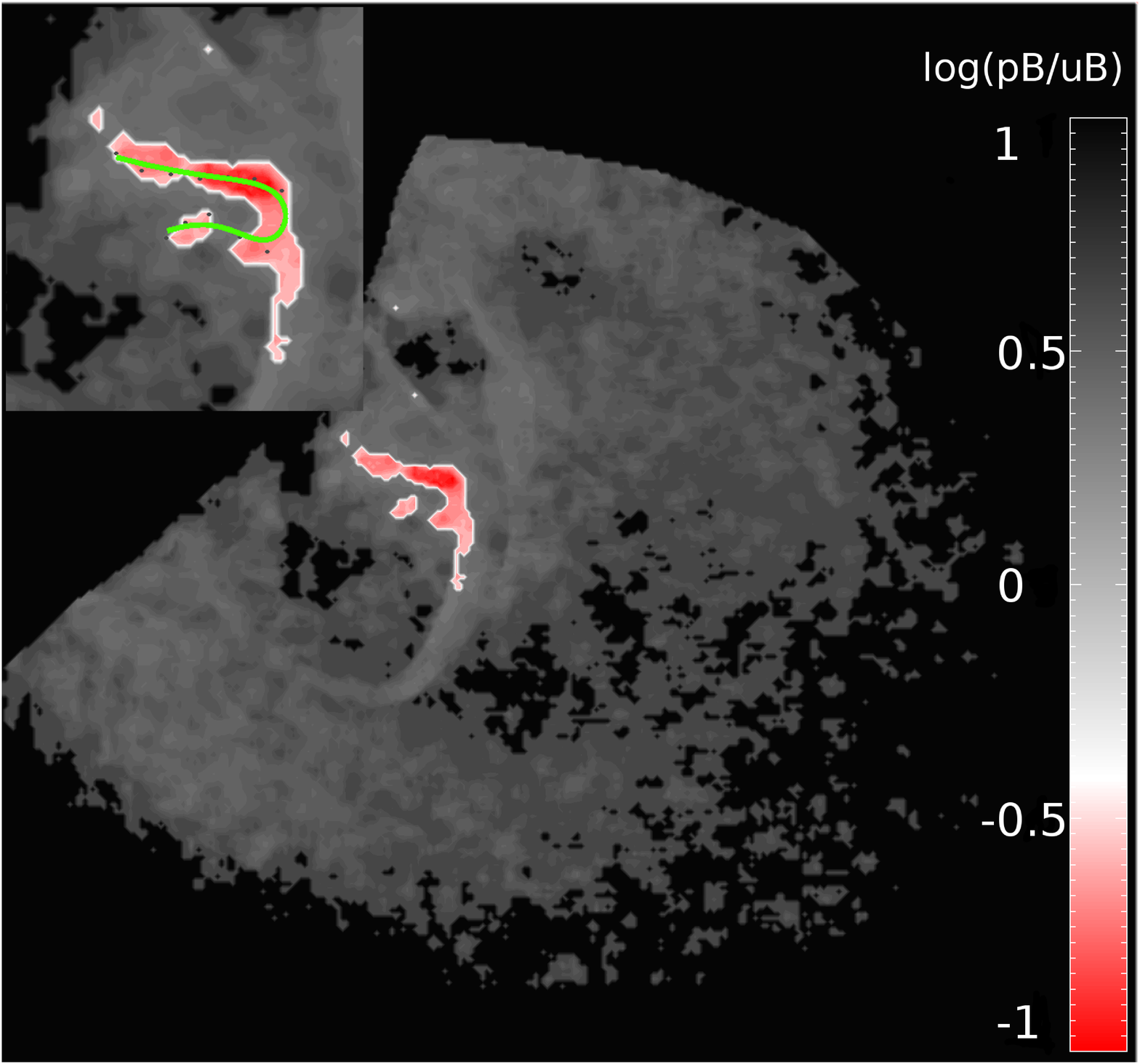}
    \\[2mm]
    \includegraphics[width=.48\textwidth,type=eps,ext=.eps,read=.eps]{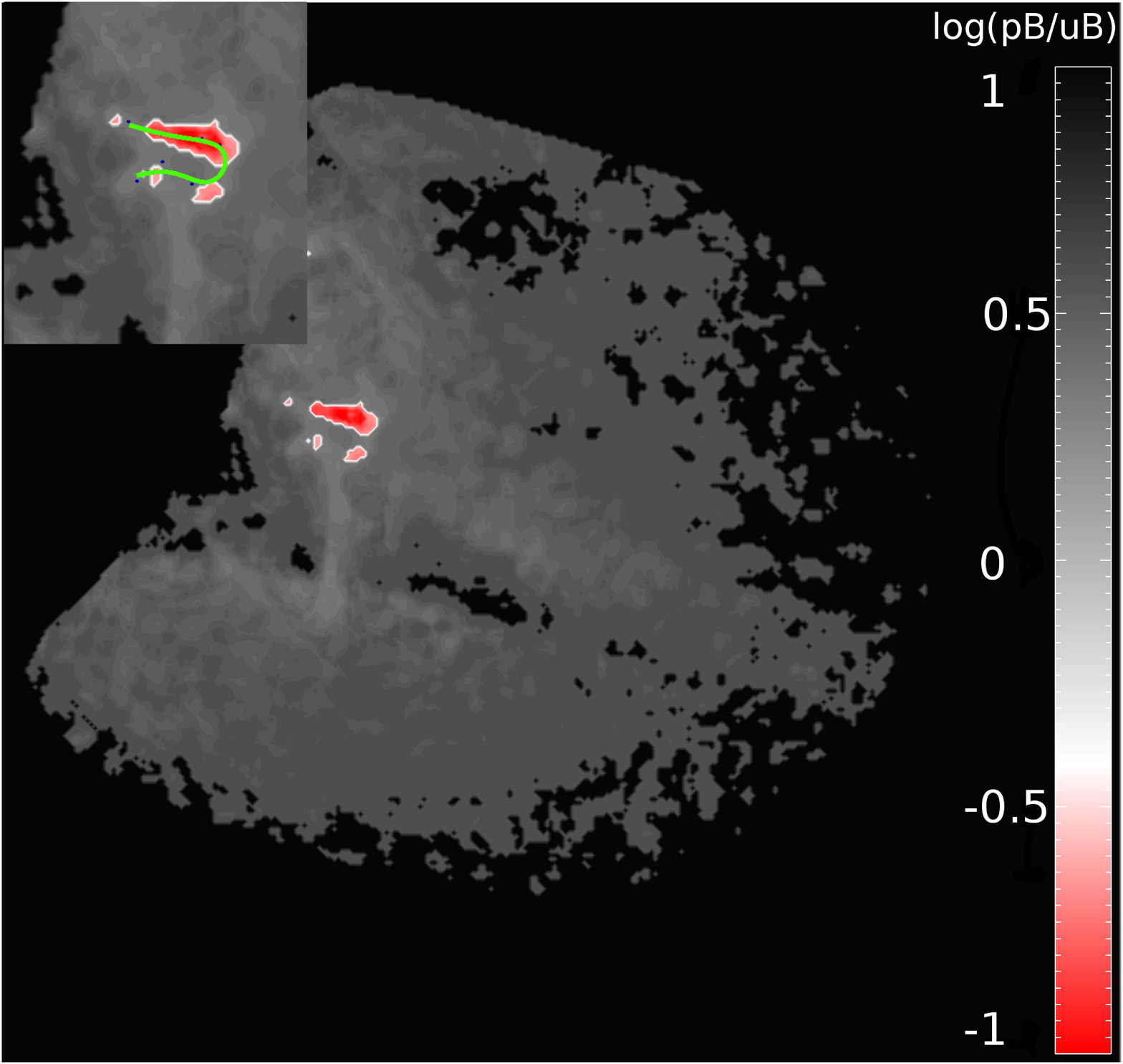} 
    \caption{Ratio $pB/uB$ of polarised to unpolarised light from
      COR1 of STEREO spacecraft A (upper panel) and  B (lower panel).
      The ratio is colour-coded and the red patches mark low polarised
      values. The upper left insert is a zoom of this region with the green line representing the projection of the 3D curve fit to the patches location obtained from stereoscopic triangulation.}
   
    \label{plotpolratio}
    \end{figure}

For a more quantitative analysis we removed the background
    contribution by subtracting a minimum intensity $pB$ and $tB$
    image from each $pB$ and $tB$ images, respectively.
    The minimum images were obtained over a time range of 12
    hours, centred at the launch time of the CME. As a result,
    we obtain the brightness of the CME alone.
In Fig.~\ref{plotpolratio} we display the resulting ratio $pB/uB$ of
polarised to unpolarised image intensities. 
From both instruments we see bright patches of extremely low polarisation
(red, $pB/tB$ $\approx$ 0.1) at about 1.5 R$_\odot$ inside the strongly polarised CME cloud (grey, $pB/tB$ $\approx$ 0.5).

The degree of polarisation of Thomson-scattered light by coronal electrons is
a function of the scattering angle between the incident light direction and
the direction towards the observer (Billings, 1966). A scattering location
close to the POS has a 90 degree scattering angle and yields a
high polarisation parallel to the limb. A scattering location far off
the POS corresponds to nearly forward/backward scattering,
which is hardly polarised. This effect allows us to estimate an effective
scattering angle from the observed degree of polarisation for each pixel
that can be related to an effective distance of the scattering location
from the POS.

Moran and Davila (2004) and Dere, Wang, and Howard (2005)
    introduced this method, called polarisation ratio method
    (PR), to construct the azimuthal barycentre plane of a CME.
    We used this technique to infer the propagation
    direction of the CME discussed here (Mierla et al. 2009).
The method is applied here to polarisation data from COR1/STEREO A and B
taken at around 21:30 UT. At each pixel of the images, the ratio
$pB/uB$ is calculated and converted into the effective scattering distance
along the line-of-sight from the POS. Owing to the forward/backward
symmetry of Thomson scattering, the brightness ratio does not indicate
whether the scatterer is in front of or behind the POS. This ambiguity
can be resolved, however, by matching the scattering locations derived
from STEREO A and B observations.

    \begin{figure}
    \centering
    \includegraphics[width=.48\textwidth,type=eps,ext=.eps,read=.eps,bb=78 100
    430 401, clip]{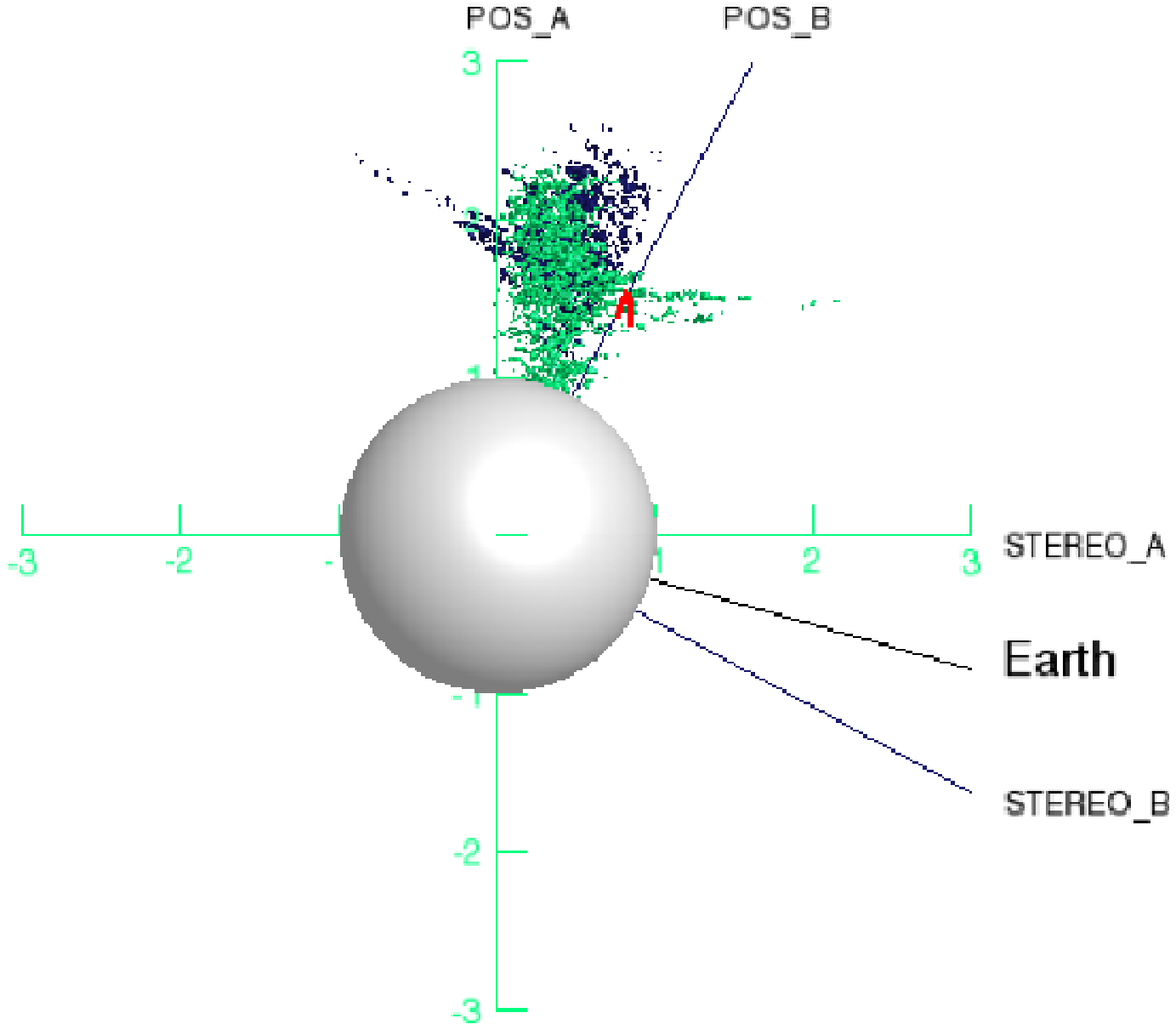}
    \caption{Polarisation ratio reconstruction of the barycentre plane of the
     CME on 31 August 2007, 21:30 UT (see section 3.1 for details).}
    \label{plotreconstr}
    \end{figure}

In Fig.~\ref{plotreconstr} we display the result of this analysis as viewed from a direction
along the northward normal of the STEREO mission plane. The green and blue dots
represent the scattering location derived from COR1/STEREO A and B pixels
of the CME cloud, respectively.
Obviously, they match well if
the barycentre of the CME is assumed in front of the
POS of STEREO A but behind the POS of STEREO B.
The only exception are the ''horns'' that extend to the far right for
STEREO A and to the far left for STEREO B pixels. These ``horns'' result from the low-polarisation patches observed in each of the COR1 images. If they were produced by Thomson scattering, the low-polarisation patch would have to be located at the tip of the "horns" in Fig.3, i.e., far away from the centre of the CME. However, these positions from STEREO A and B are several solar radii apart and are therefore inconsistent.

 An independent estimate of the location of the low-polarisation patch can be obtained by stereoscopy, provided we assume that the patches we see in both coronagraphs are the same object. Both in brightness (see Fig.~\ref{plotsmin11aug}) and in polarisation ratio
(see Fig.~\ref{plotpolratio}) the patches displayed a high
contrast and a sharp boundary with respect to the CME environment.
The COR1 instrument of STEREO B seems somewhat less sensitive because
not all features of the STEREO A patch can be matched to an equivalent
signal in the STEREO B image. Yet we could determine a loop-like 3D curve,
the projections of which trace out the patches in COR1/STEREO B and the
major structures in COR1/STEREO A. These projections are drawn in green
into the inserts of Fig.~\ref{plotpolratio}.

The 3D curve is again projected into Fig.~\ref{plotreconstr}, in red. The triangulation of the patches locates it very close to the barycentre plane of the CME. Note that the cloud in Fig.~\ref{plotreconstr} does not show the full extent of the CME, but is only an approximation of its azimuthal barycentre plane. The PR method cannot resolve the azimuthal extent of a CME (see Mierla et al., 2011). 

\section{Discussion}

Low polarisation in coronagraph images can have several explanations:
1) F-corona emission (\eg, Morgan and Habbal 2007),
2) Thomson scattering from enhanced plasma density far away from the POS
   (\eg, Billings 1966), and
3) H$\alpha$ emission (\eg, Poland and Munro, 1976).
F-corona contributions can be ruled out because it forms a diffuse
background and does not vary rapidly in time. The subtraction of the
minimum background intensities from our images should have removed any
F-corona contribution.
The interpretation of the low-polarisation patches by Thomson
scattering can also be ruled out. We showed in the previous
section that this assumption leads to different locations of the
scatterer for STEREO A and B observations, which are both
inconsistent with the results from the stereoscopic triangulation.

We are therefore left with the last explanation. H$\alpha$ patches in
white-light coronagraph images have only
rarely been reported, even though bright amorphous structures are often seen in CME clouds. Coronal mass ejections often appear in coronagraph images
as three-part structures composed of a bright leading edge, a dark cavity, and
a bright core, which are associated with the compressed solar plasma ahead of the
ejecta, the erupting magnetic flux rope, and the cool and dense prominence
plasma, respectively (\eg, Cremades and Bothmer 2004). However, a
convincing identification of the white-light core of the CME with the cool
prominence material is relatively rare.
Even if recent studies have demonstrated a strong connection between 
prominence eruptions and CMEs, the correlation is not always one-to-one.

It is not quite clear how much of the emission from the bright core
structures are produced by Thomson scatter from its enhanced plasma
density and how much stems from H$\alpha$ radiation. Because we observed
the patches at an outward velocity above 100 km/s, their H$\alpha$ resonance
is well outside of the H$\alpha$ absorption line from the solar surface
spectrum. Even though the depth of the absorption is 1/7 of the surrounding
continuum, the H$\alpha$ radiation emitted from the rising patch
is only little enhanced by this Doppler brightening effect because of
the complicated balance of the electronic level population of hydrogen
atoms exposed to the solar spectrum (\eg, Hyder and Lites 1970).

The observed brightness $B_\mathrm{patch}$ of the H$\alpha$ patch in
the image is in fact a line-of-sight (LOS) superposition of the
radiation from the three different sources: the H$\alpha$ radiation,
$B_{\mathrm{H}\alpha}$, and the Thomson scatter, $B_\mathrm{Th'}$,
from inside the H$\alpha$ cloud and the ambient Thomson scatter,
$B_\mathrm{Th}$, along remaining part of the LOS through the patch.
The latter contribution can roughly be assumed equal to the Thomson
scatter measured close to, but outside the H$\alpha$ patch.
  
Both the total and the polarised intensities add (assuming that the 
polarisation direction in all three components is the same), 
\begin{gather}
tB_\mathrm{patch}=tB_{\mathrm{H}\alpha}+tB_\mathrm{Th'}+tB_\mathrm{Th}
\label{tBpatch}\\
pB_\mathrm{patch}=pB_{\mathrm{H}\alpha}+pB_\mathrm{Th'}+pB_\mathrm{Th}.
\label{pBpatch}
\end{gather}
We find that the total brightness of the H$\alpha$ patch (around 1--5$\cdot10^{-8}$ MSB, where MSB is mean solar brightness) is about 
10 times as high as a typical value in the Thomson-scattering area (around 1--5$\cdot10^{-9}$ MSB), i.e.,
$tB_\mathrm{patch} \simeq 10\;tB_{Th}$. 
According to the Thomson-scattering theory (Billings 1966), the calibrated value of $tB_\mathrm{Th}$ corresponds to an electron column density along the LOS of 1.8 10$^{17}$ cm$^{-2}$.
For a depth of the CME cloud of about 1 R$_\odot$, this yields 
an excess density of the CME over the streamer background of
2.6 10$^6$ cm$^{-3}$.

For the polarisation ratio we find (see Fig.~\ref{plotpolratio})
\begin{equation}
 r=\frac{pB}{tB}=\frac{\frac{pB}{uB}}{1+\frac{pB}{uB}}\simeq
\begin{cases}
  0.5 & \text{for}\; r_\mathrm{Th}\\
  0.1 & \text{for}\; r_\mathrm{patch},
\end{cases}
\nonumber
\end{equation}
where $r=pB/tB$ is the polarisation ratio of the respective component.
For Thomson scatter, $r_\mathrm{Th}$ should depend only on the distance from
the solar surface, hence we can assume $r_\mathrm{Th} \simeq r_\mathrm{Th'}$.
Obviously, $tB_{\mathrm{H}\alpha}$ must be significant in (\ref{tBpatch})
over the Thompson scatter contribution because $r_\mathrm{patch}$ 
differs considerably from $r_\mathrm{Th}$.

Inserting the observed total brightness and polarisation ratios into
(\ref{tBpatch}) and (\ref{pBpatch})
and eliminating $tB_\mathrm{Th}$, we obtain the
relation
\begin{equation}
\frac{tB_{\mathrm{H}\alpha}}{tB_\mathrm{Th'}}=
\frac{8}{1-18\,r_{\mathrm{H}\alpha}}.
\nonumber
\end{equation}
Judging from the brightness and polarisation ratios observed,
$r_{\mathrm{H}\alpha}$ cannot be higher than 1/18. This
low value of the intrinsic H$\alpha$ polarisation ratio agrees
with the low values obtained for chromospheric measurements
(Gandorfer, 2000; Wiehr and Bianda, 2003).
Moreover, the ratio $tB_{\mathrm{H}\alpha}/tB_{Th'}$
cannot be lower than 8. Consequently a large part of the radiation from
the core patch (at least 88\%) must be H$\alpha$ emission.

Jej\v{c}i\v{c} and Heinzel (2009) have calculated 
the ratio $tB_{\mathrm{H}\alpha}/tB_{Th'}$ 
for various temperatures and  densities 
(note that they assume a different white-light 
bandwidth 10 nm instead of 22.5 nm and a dilution factor $W$=0.416 
instead of 0.21 appropriate for 1.5 R$_\odot$. This causes our white-light intensities to be 1.3 times brighter than the model intensities
assumed in their paper. We neglect this factor in view of the approximate
nature of our estimate).
For the comparatively low densities we expect in the corona,
they derive a linear relation, 
$tB_{\mathrm{H}\alpha}/tB_\mathrm{Th'}=n_e/10^8 \mathrm{cm}^{-3}$,
which only weakly depends on temperature, provided it is below
15000~K. If these conditions hold in our case, the density in the
H$\alpha$ patch would exceed 8 10$^8$ cm$^{-3}$ and would hence be
nearly three orders of magnitude of what we estimated for the CME
cloud outside of the patch.

The brightness from the core patch in the CME is dominated
     by H$\alpha$ radiation. If, reversely, we had assumed
     $tB_\mathrm{patch}$ to be entirely caused by Thomson scatter, we
     would have obtained a gross overestimate of the density in the
     CME core. In view of this result, some CME mass estimates in
     previous studies, where the contribution of the core brightness
     was not negligible, may have to be revised.

\section{Conclusions}

We showed that the white-light core of the 31 August 2007 CME is clearly identified with the eruptive prominence observed in the EUVI 304 images. To our knowledge, we demonstrate for the first time that this core material is located close to the centre of CME cloud. Moreover, we showed that the major part of the CME core emission, more than 85\% in our case, is H$\alpha$ radiation and only a small fraction is Thomson-scattered light. We made a rough estimate of the electron density, showing that the density in the H$\alpha$ patch will exceed by nearly three orders of magnitude what we estimated for the CME cloud outside of the patch.

\begin{acknowledgements} 
MM would like to thank T. Moran and A. Vourlidas for constructive discussions on PR method and electron densities.
The authors acknowledge the use of SECCHI data.
The contribution of the IC and BI benefited form support of the
German Space Agency DLR and the German ministry of economy and
technology under contract 50 OC 0904.
BI thanks A. Gandorfer for enlightening discussions on the nature 
of the H$\alpha$ radiation.
\end{acknowledgements}

\end{document}